\documentclass[manuscript]{acmart}

\AtBeginDocument{%
  \providecommand\BibTeX{{%
    \normalfont B\kern-0.5em{\scshape i\kern-0.25em b}\kern-0.8em\TeX}}}

\setcopyright{acmlicensed}
\copyrightyear{2025}
\acmYear{2025}
\acmDOI{XXXXXXX.XXXXXXX}
\usepackage{array}

\usepackage{tabularx}
\usepackage{multirow}

\begin{document}

\title[From Reports to Reality]{From Reports to Reality: Testing Consistency in Instagram’s Digital Services Act Compliance Data}

\author{Marie-Therese Sekwenz}
\affiliation{%
  \institution{Delft University of Technology}
  \city{Delft}
  \country{Netherlands}
}
\email{m.t.sekwenz@tudelft.nl}

\author{Ben Wagner}
\affiliation{%
  \institution{I:TU Interdisciplinary Transformation University Austria, Delft University of Technology, Inholland}
  \city{Delft, Linz}
  \country{Netherlands, Austria}}
\email{B.Wagner@tudelft.nl}

\author{Hans de Bruijn}
\affiliation{%
  \institution{Delft University of Technology}
  \city{Delft}
  \country{Netherlands}
}
\email{J.A.deBruijn@tudelft.nl}

\renewcommand{\shortauthors}{Anonymous et al.}

\begin{abstract}
The Digital Services Act (DSA) introduces harmonized rules for content moderation and platform governance in the European Union, mandating robust compliance mechanisms, particularly for very large online platforms and search engines. This study examined compliance with DSA requirements, focusing on Instagram as a case study. We develop and apply a multi-level consistency framework to evaluate DSA compliance. Our findings contribute to the broader discussion on empirically-based regulation, providing insight into how researchers, regulators, auditors and  platforms can better utilize DSA mechanisms to improve reporting and enforcement quality and accountability. This work underscores that consistency can help detect potential compliance failures. It also demonstrates that platforms should be evaluated as part of an interconnected ecosystem rather than through isolated processes, which is crucial for effective compliance evaluation under the DSA.
\end{abstract}




\maketitle

\section{Introduction}
\label{Introduction}
The Digital Services Act (DSA) introduces a comprehensive regulatory framework for platform governance in the European Union, particularly targeting very large online platforms (VLOPs) and very large online search engines (VLOSEs) (Art. 33) \cite{noauthor_regulation_2022}.

The DSA builds on lessons from earlier national regulations (see Section \ref{Related}). Despite these advances, previous studies have shown that transparency alone does not ensure accountability \cite{heldt_reading_2019,parsons_effectiveness_2019,wagner_regulating_2020}. The DSA instead enables examination of platform governance through an integrated regulatory ecosystem, supporting more meaningful accountability than previous isolated indicators allowed.

Through a suite of new compliance obligations—including transparency reports (Arts. 15, 24, 42), statements of reason (SOR) in the DSA Transparency Database (Arts. 17, 24(5)) that document all content moderation decisions taken by platforms, systemic risk assessments (Art. 34) evaluating platforms internally, independent audits (Art. 37) cross-checking systemic risk assessments by third parties, and operating advertising repositories to collect information on political content (Art. 39)—the DSA seeks to offer a multitude of new mechanisms of control to evaluate online platforms.

We use \emph{consistency} as a method for evaluating platform compliance, focusing on Instagram. Building on Keller’s work, we treat reporting inconsistencies as possible indicators of misreporting or governance failure \cite[p. 18]{keller_discussion_1968, keller_compliant_2023, keller_dsa_2022}. In the financial domain, consistency provides a materiality threshold that allows auditors to assess whether reported activities can be trusted \cite{koutsos_cross_2024}. Analogously, in platform governance, consistency enables evaluators to identify discrepancies between mechanisms—such as mismatches between reported account suspensions in transparency reports and moderation actions recorded in the SOR (See Section \ref{External consistency}).

In this paper, we assess how different forms of consistency can be used to evaluate compliance with the DSA by analyzing three of its key transparency and oversight mechanisms: (1) biannual transparency reports (number 1-4), (2) statements of reason (SOR) published in the Transparency Database, (3) the first systemic risk assessment and (4) independent audit reports. Focusing on Instagram as a case study, we ask:
\begin{itemize}
\item \textbf{RQ1:} How can different forms of consistency be assessed across DSA reporting instruments?
\item \textbf{RQ2:} What challenges and opportunities arise in applying consistency checks to a VLOP’s DSA compliance?
\item \textbf{RQ3:} How can a consistency-based approach support effective oversight and auditing under the DSA?
\end{itemize}

Our contributions are threefold. First, we propose a framework for multi-level consistency assessment of DSA compliance, informed by audit theory, legal scholarship, and empirical research. Second, we apply this framework to Instagram’s transparency reports, SOR data, and risk assessments. Third, we identify practical implications for researchers, NGOs, regulators, and auditors.

The rest of the paper is structured as follows: Section 2 reviews related work and legal foundations, Section 3 presents our methodology, Section 4 applies our consistency framework to Instagram, Section 5 discusses key challenges, Section 6 outlines opportunities and recommendations, and Section 7 concludes.

Our findings inform ongoing discussions about regulatory implementation, compliance evaluation, and auditability under the DSA. By bridging granular decision-level data (from SORs) with aggregate reporting (from transparency reports) and risk-based assessments (from systemic risk evaluations), we provide a structured approach to understanding how consistency can be used as a tool for empirical compliance assessment and regulatory scrutiny.

\section{Methodology}
\label{Methodology}
This study develops and applies a consistency-based methodology to evaluate how control mechanisms introduced by the Digital Services Act (DSA) can be used to assess platform compliance. 

Focusing on Instagram — a very large online platform (VLOP) with high regulatory obligations according to Article 33 DSA — we analyze how these four key mechanisms (defined in Section \ref{Introduction}) operate in practice for Instagram. Together, these mechanisms form a layered system of regulatory oversight that enables both qualitative and quantitative compliance evaluations.

To assess how these instruments function as tools of control, we propose and apply a multi-level consistency framework, distinguishing between:
\begin{itemize}
\item \textbf{Internal consistency} — alignment within a single report or mechanism;
\item \textbf{External consistency} — coherence across multiple reporting instruments (e.g., between SORs, transparency reports, and risk assessments);
\item \textbf{Historical consistency} — comparability across reporting periods; and
%\item \textbf{(Later added)} Cross-platform consistency — comparability across VLOPs and VLOSEs (see Section \ref{RQ4}).
\end{itemize}

Our methodological approach combines:
\begin{enumerate}
\item A doctrinal legal analysis to identify the obligations, scope, and interrelations of the relevant DSA mechanisms, based on primary legal texts \cite{eu2024transparencyregulation,EU2023DSADelegatedAudit}.
\item A compliance evaluation of Instagram’s reporting practices, drawing from its publicly submitted DSA transparency reports (Rounds 1–4 \cite{meta2023dsa1,meta2024dsa2, meta2024dsa3, meta2025dsa4}), its SOR entries from the DSA Transparency Database (capturing over 32 million moderation decisions in May 2025 \cite{european_commission_commission_2023}), and the first available systemic risk assessment and independent audit reports \cite{meta_sra_2024, meta_audit_2024}.
\item A comparative consistency check to triangulate data across these mechanisms, enabling us to assess the plausibility, coherence, and completeness of Instagram’s reporting under the DSA framework.
\end{enumerate}

The relevance of consistency directly maps onto our research questions as introduced earlier (see Section \ref{Methodology}). Together, these questions reframe DSA reporting not as discrete compliance actions, but as a system of interlocking mechanisms whose integrity hinges on their internal and mutual coherence.

Whereas previous studies have documented inconsistencies in single reporting streams, this paper uniquely triangulates four DSA mechanisms using a multi-level consistency framework, and demonstrates cross-mechanism gaps in a case study.

By synthesizing findings from all four mechanisms, our methodology enables a holistic evaluation of whether Instagram’s reporting aligns with the DSA’s control logic. It also highlights where reporting practices fall short of delivering accountability. In doing so, this paper contributes both an analytical lens and a practical toolset for regulators, auditors, and researchers tasked with interpreting the increasingly complex landscape of platform compliance under the DSA.

\subsection{Case study selection: Instagram}
We selected Instagram as a case study due to its significance as a VLOP with an estimated 272.0 million monthly active users in the EU \cite{meta2025dsa4}, its ongoing regulatory scrutiny, and the public availability of its submissions across the three examined mechanisms. 

The European Commission has launched formal proceedings into Instagram's compliance with the DSA, including concerns over political content visibility, deceptive advertising, and takedown mechanisms \cite{noauthor_commission_2024-1__open_Meta} — making it a timely and relevant case for identifying potential inconsistencies and accountability gaps. Similarly, in May 2025, the Commission concluded that TikTok had breached its obligations under the DSA regarding advertising transparency, making the first steps toward DSA enforcement and underscoring the broader challenges of ensuring compliance across major platforms \cite{EU2025TikTokDSA}.

For the SOR data, we analyzed moderation decisions submitted during the periods covered by Transparency Reports 2-4, starting from the operational launch of the SOR database on 26 September 2023 \cite{noauthor_digital_2024}. In total, Instagram reported over 58 million moderation actions during these periods \cite{noauthor_commission_2023}. For systemic risk assessments, we examined Instagram’s first published systemic risk report in line with Article 34 DSA, focusing on its structure, self-reported mitigation strategies (Art. 35), and correspondence with transparency and SOR moderation data. The independent audit reports conducted by third party auditors like KPMG, FTI Consulting, Holistic AI, or EY are regulated under Article 37 DSA and act as a control mechanism for the systemic risk assessment reports the platforms issue. Similar as for Article 34 reports we analyse the first issed round of audit reports for Instagram \cite{meta_audit_2024}.

\section{Related Work}
\label{Related}

Mechanisms for platform accountability predate the Digital Services Act (DSA), with early legislative frameworks such as Germany’s Network Enforcement Act (NetzDG) and Austria’s Communications Platform Law (KoPl-G) mandating transparency reports from online platforms \cite{noauthor_german_2017, noauthor_bundesgesetz_2020}. These laws required semi-annual disclosures of content moderation activities, but were widely criticized for producing non-standardized, opaque reporting structures and category inconsistencies \cite{wagner_regulating_2020, heldt_reading_2019, zipursky_nuts_2019}. For instance, Facebook was fined €9 million under NetzDG for inadequate reporting \cite{noauthor_germany_2019}.

Alongside these legal obligations, voluntary transparency efforts such as Wikimedia's 2017 report \cite{buatti_wikimedia_2017}, the Santa Clara Principles (2018), and the OECD Voluntary Transparency Framework (2020) aimed to enhance platform accountability, yet lacked enforceability and methodological coherence \cite{parsons_effectiveness_2019, aclu_santa_nodate, oecd_current_2020}.
These early initiatives demonstrated both the potential and limitations of transparency mechanisms in platform governance.

The DSA builds on this background, introducing an enforceable, layered regime of control mechanisms. These include enhanced transparency reports (Articles 15, 24, 42), Statement of Reasons (SOR) databases (Articles 17, 24(5)), systemic risk assessments (Article 34), and independent audits (Article 37), all further harmonized by the Delegated Regulation (DR) under Articles 24 and 88 DSA \cite{noauthor_regulation_2022, EU2023DSADelegatedAudit, eu2024transparencyregulation}. The DSA’s legal requirements operationalize compliance monitoring, ensure traceability of decisions, and provide regulators with audit-capable documentation.

The DSA introduces tiered reporting obligations. Article 15 sets basic transparency requirements for intermediary services, mandating machine-readable, publicly accessible reports. Article 24 extends these requirements to platforms, and Article 42 imposes comprehensive biannual reporting for very large online platforms (VLOPs) and search engines (VLOSEs). Transparency reports must detail notice-and-action mechanisms (Article 16), the use of automated tools (Article 15(1)(b)), human moderation (Article 15(1)(e)), and provide a breakdown of moderation actions by legal or contractual grounds (Article 15(1)(c)). The Statement of Reasons (Article 17) database further requires platforms to publish justifications for each moderation action, including the legal or contractual basis.

Platforms’ contractual governance is also addressed, with Article 14 anticipating a Terms and Conditions (TaC) Database for centralizing platform rules and enhancing transparency \cite{noauthor_commission_2023}. Article 34 introduces systemic risk assessments, requiring VLOPs/VLOSEs to identify, analyze, and mitigate systemic risks, e.g. impacting fundamental rights, imposing negative effects on the electoral process, or dissiminating illegal content. These assessments are subject to independent audits under Article 37 and further detailed in the Delegated Regulation (DR), which standardizes reporting and auditing, risks, and sets standards for methodologies \cite{EU2023DSADelegatedAudit}.

Scholars have conceptualized traceability, auditability, and verifiability as key pillars of trustworthy algorithmic systems. For example, Kroll et al.\ highlight procedural regularity as foundational for meaningful oversight \cite{kroll2017accountable}, while Binns et al.\ warn that transparency without consistency risks masking deeper systemic opacity \cite{binns2018fairness}. Thus, consistency is not merely a technical goal but a structural precondition for effective compliance monitoring.

Recent empirical studies of DSA implementation illustrate both advances and persistent challenges. Trujillo et al.\ analysed 195 million SOR entries (September 2023–January 2024), revealing a 93\% automated detection rate for Instagram’s moderation and highlighting ambiguities in how automated tools were documented \cite{trujillo_dsa_2024}. Drolsbach and Pröllochs studied Instagram’s early DSA months, identifying platform-specific moderation differences \cite{drolsbach_content_2023}. Kaushal et al.\ critiqued vague justification references to platforms’ terms and conditions (TaC) \cite{kaushal_automated_2024}. 
Terzis has further advanced the field by empirically investigating the practices and limitations of algorithmic audits under the DSA, revealing the need for more robust audit methodologies and independent oversight \cite{terzis_law_2024}.
These findings underscore both the analytical richness of new DSA data and the urgent need for interpretive, consistency-based tools.

In sum, the DSA and its delegated acts establish a multi-layered framework for regulatory control, positioning consistency as a central pillar for effective oversight. Discrepancies across mechanisms—internal, external, or historical—signal weaknesses in platform governance and invite regulatory scrutiny.

\section{Three Levels of Consistency: A New Methodology to Evaluate the Compliance of DSA reporting mechanisms}
\label{RQ2}

\begin{figure}[ht]
\centering
\includegraphics[width=0.8\textwidth]{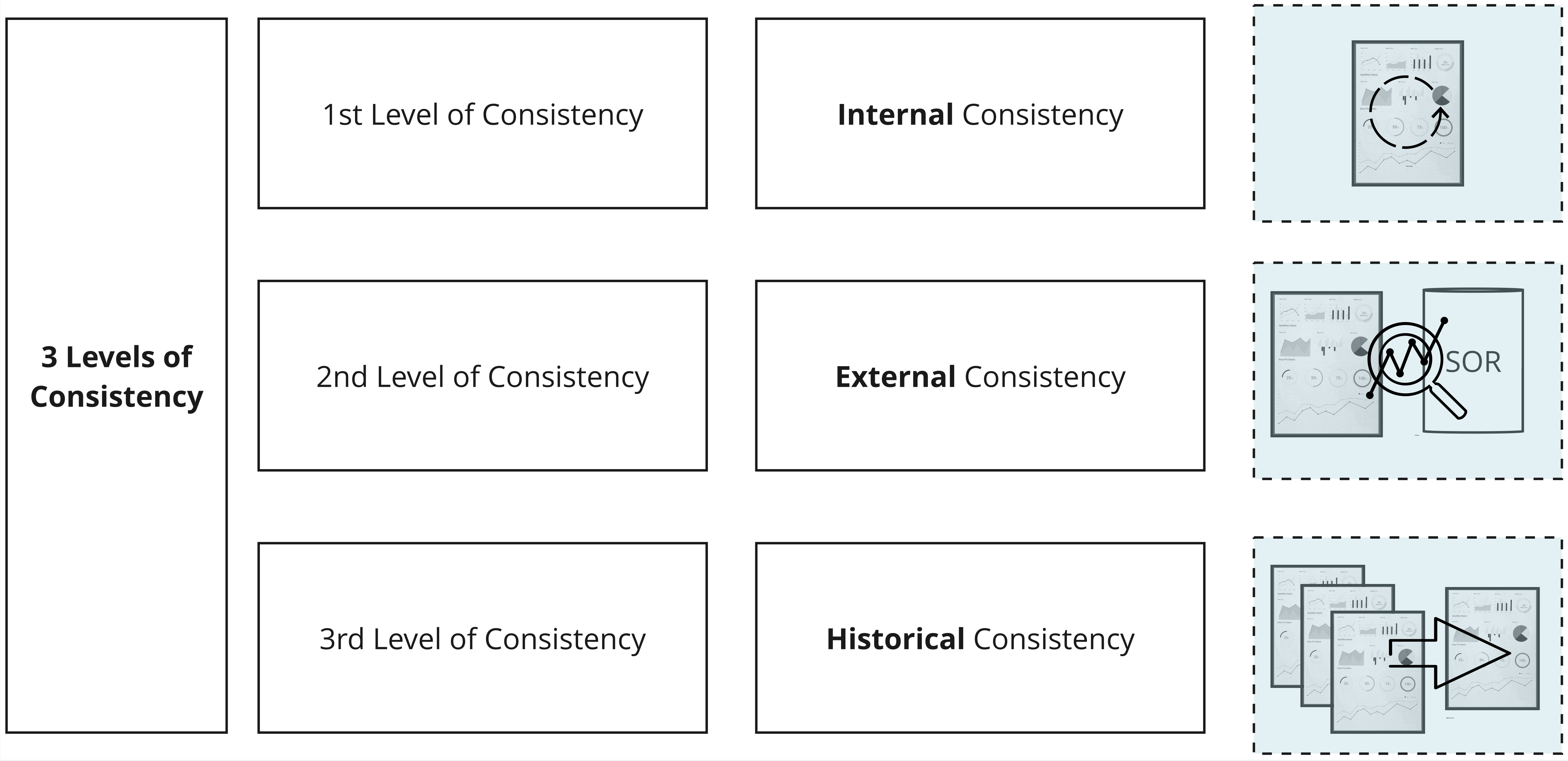}
\caption{Three levels of consistency for evaluating transparency reporting and the information provided in the DSA Transparency Database (SOR).}
\Description{Diagram showing three analytical levels of consistency: (1) within a single reporting mechanism, (2) across different reporting mechanisms such as transparency reports and the Statement of Reasons database, and (3) over time across multiple reporting periods. Used to evaluate reliability and comparability of DSA mechanism disclosures.}
\label{fig:3level}
\end{figure}

In this section, we answer RQ1: \emph{What methodology can be used to evaluate different forms of consistency of the means of transparency within the scope of the DSA for a VLOP?} Therefore, we propose three levels of consistency, as illustrated in \ref{fig:3level}, for evaluating transparency reporting and the information provided in the DSA Transparency Database for testing compliance under the DSA:
\begin{itemize}
    \item \textbf{Level 1}: Internal consistency: This ensures internal consistency in reporting, controlling for inconsistencies within the report under investigation by, for example, checking sums within a table.
    \item \textbf{Level 2}: External consistency: This tests reporting for consistency across two or more reporting mechanisms by, for example, comparing the total number of suspensions of the service from a transparency report with the correlating number of SORs with the same reason.
      \item \textbf{level 3}: Historical consistency: This ensures consistency throughout reporting time frames. Information from previous reports can be compared with new reporting data or information. One example of a historical comparison is the comparison the average reported numbers of service recipients between year X1 and X2.
\end{itemize}

\subsection{Internal Consistency}
\label{Internal consistency}

We assess internal consistency by comparing reported totals with subtotals in Instagram’s transparency reports. In several cases (e.g., Reports 1–4), the sum of disaggregated content removal entries does not match the reported totals, with deviations of 71,400,270 (Report 1), 1,370,980 (Report 2), 1,253,173 (Report 3), and 147,746 (Report 4) instances, respectively. This suggests ambiguity in the 'Total (including other violations)' category, which lacks clear legal or contractual definition by only providing a lump sum.

\begin{table*}[ht]
\centering
\caption{Instagram Content Removals by Policy Category Across DSA Transparency Reports 1–4 (Total and Automated Removals)}
\label{tab:instagram_combined_removals}
\resizebox{\textwidth}{!}{%
\begin{tabular}{>{\raggedright\arraybackslash}p{6cm}|r r|r r|r r|r r}
\toprule
\textbf{Policy Category} & \multicolumn{2}{c|}{\textbf{Report 1}} & \multicolumn{2}{c|}{\textbf{Report 2}} & \multicolumn{2}{c|}{\textbf{Report 3}} & \multicolumn{2}{c}{\textbf{Report 4}} \\
 & Total & Automated & Total & Automated & Total & Automated & Total & Automated \\
\midrule
Adult Nudity and Sexual Activity & 950,816 & 852,471 & 1,150,038 & 1,057,008 & 1,351,522 & 1,213,764 & 898,264 & 778,404 \\
Adult Sexual Solicitation / Explicit Language & -- & -- & -- & -- & -- & -- & 97,855 & 44,416 \\
Bullying and Harassment & 837,741 & 722,502 & 1,297,196 & 1,131,702 & 1,176,634 & 1,015,909 & 480,721 & 329,539 \\
Child Endangerment / Nudity & 47,154 & 14,077 & 22,582 & 7,832 & 133,229 & 5,927 & 84,166 & 22,760 \\
Child Sexual Exploitation & 60,364 & 16,886 & 145,984 & 107,142 & 110,226 & 82,605 & -- & -- \\
Cybersecurity & -- & -- & -- & -- & -- & -- & 3,406 & 3,327 \\
Dangerous Orgs – Hate Orgs & 46,338 & 23,433 & 70,430 & 41,212 & 45,839 & 26,225 & -- & -- \\
Dangerous Orgs – Terrorism & 26,038 & 21,571 & 166,098 & 151,789 & 76,224 & 66,356 & -- & -- \\
Dangerous Individuals and Orgs & -- & -- & -- & -- & -- & -- & 100,862 & 79,908 \\
Fraud and Deception & -- & -- & -- & -- & -- & -- & 115,444 & 32,603 \\
Hate Speech & 1,521,669 & 1,352,204 & 1,509,400 & 1,390,887 & 1,450,002 & 1,331,715 & 716,246 & 639,852 \\
Restricted Goods – Drugs & 49,478 & 44,405 & 45,810 & 40,893 & 32,877 & 22,290 & -- & -- \\
Restricted Goods – Firearms & 7,094 & 5,398 & 33,806 & 32,150 & 11,571 & 10,858 & -- & -- \\
Restricted Goods and Services (Merged) & -- & -- & -- & -- & -- & -- & 145,910 & 24,538 \\
Spam & -- & -- & 29,628,165 & 29,622,615 & 5,200,240 & 5,076,296 & 3,242,761 & 3,236,558 \\
Suicide, Self-Injury & 153,051 & 121,530 & 92,854 & 78,994 & 106,042 & 71,835 & -- & -- \\
Suicide, Self-Injury and Eating Disorders & -- & -- & -- & -- & -- & -- & 51,638 & 35,805 \\
Third-Party Intellectual Property Infringement & -- & -- & -- & -- & -- & -- & 145,848 & 41,725 \\
Violence and Incitement & 1,188,216 & 979,568 & 1,400,781 & 1,297,277 & 1,033,996 & 949,394 & 556,201 & 443,079 \\
Violent and Graphic Content & 10,184 & 8,020 & 105,287 & 102,390 & 155,372 & 146,248 & 28,833 & 24,993 \\
\textbf{Total (incl. other violations)} & \textbf{76,298,413} & \textbf{75,113,462} & \textbf{37,039,411} & \textbf{35,724,613} & \textbf{12,136,947} & \textbf{10,535,490} & \textbf{6,760,857} & \textbf{5,800,826} \\
\bottomrule
\end{tabular}%
}
\end{table*}

Such aggregation practices contradict Annex II of the Delegated Regulation, which mandates that totals must represent the sum of disaggregated categories and that double-counting should be omitted \cite[p. 7]{noauthor_annex_2024}. Furthermore, inconsistencies arise from how Instagram presents cross-platform figures. For example, the 'Number of business content removal measures' aggregates Instagram and Facebook data, while adjacent tables refer exclusively to Instagram (see Table 15.1.c.(3) \cite{meta2025dsa4}). This blending obfuscates accountability and violates principles of platform-specific traceability as outlined in Article 15(1)(c) of the DSA as "meaningful and comprehensible information" for content moderation.

In terms of illegal content reporting under Article 16, broad categories such as 'Other Illegal Content' remain vague and fail to align with the 17 defined categories in Annex II of the DR \cite[p. 7-11]{noauthor_annex_2024}. Instagram’s failure to distinguish, for instance, terrorist content or hate speech within these submissions undermines legal clarity and conflicts with Article 16(6) obligations to provide granular reporting highlighted by the need for "decisions in respect of the information to which the notices relate" (see Table 15.1.b.(1) in Instagram's reports \cite{meta2025dsa4}).

Additionally, Instagram’s claim that "all Article 16 notices are processed using manual review" \cite[p. 9]{meta2025dsa4} lacks specificity. It remains unclear whether this applies to all content flagged as illegal under the DSA or only subsets. Instagram's report demotion and removal statistics further reflect inconsistency in granularity: while some categories are broken down by guideline, others remain vaguely grouped, complicating internal consistency assessments (see Table 15.1.c.(2) \cite{meta2023dsa1, meta2024dsa2, meta2024dsa3, meta2025dsa4}).

\begin{table*}[ht]
\centering
\caption{Instagram Notices Submitted Through Article 16 Mechanisms in DSA Transparency Reports 1–4 (See Table 15.1.b.(1) in the reports)}
\label{tab:instagram_illegal_expanded}
\resizebox{\textwidth}{!}{%
\begin{tabular}{|p{3.5cm}|r r r r|r r r r|}
\hline
\textbf{Type} &
\multicolumn{4}{c|}{\textbf{Notices Submitted}} &
\multicolumn{4}{c|}{\textbf{Content Actioned}} \\
& \textbf{Report 1} & \textbf{Report 2} & \textbf{Report 3} & \textbf{Report 4} & \textbf{Removed (R1)} & \textbf{Removed (R2)} & \textbf{Removed (R3)} & \textbf{Removed (R4)} \\
\hline
Intellectual Property (IP)     & 231,334   & 103,244 & 89,859 & 43,895 & 74,336 & 37,262 & 29,657 & 14,729 \\
Defamation                     & 64,966    & 111,252 & 62,929 & 35,753 & 12,138 & 20,580 & 11,953 & 8,612  \\
Privacy                        & 7,626     & 29,878  & 32,267 & 22,224 & 1,305  & 6,134  & 6,078  & 3,216  \\
Other Illegal Content          & 47,477    & 151,633 & 137,228 & 61,888 & 11,065 & 28,962 & 28,396 & 12,418 \\
\hline
\textbf{Total}                 & \textbf{351,403} & \textbf{396,007} & \textbf{322,283} & \textbf{163,760} & \textbf{98,844} & \textbf{92,938} & \textbf{76,084} & \textbf{38,975} \\
\hline
\end{tabular}%
}

\vspace{0.2cm}

\small
\begin{tabular}{|p{3.5cm}|r r r r|}
\hline
\textbf{Type} &
\multicolumn{4}{c|}{\textbf{Content Restricted}} \\
& \textbf{Report 1} & \textbf{Report 2} & \textbf{Report 3} & \textbf{Report 4} \\
\hline
Intellectual Property (IP)     & 0   & 0   & 0   & 0   \\
Defamation                     & 80  & 294 & 117 & 41  \\
Privacy                        & 4   & 15  & 11  & 4   \\
Other Illegal Content          & 310 & 2,199 & 1,774 & 655 \\
\hline
\textbf{Total}                 & \textbf{394} & \textbf{2,508} & \textbf{1,902} & \textbf{700} \\
\hline
\end{tabular}%
\end{table*}

\subsection{External Consistency}
\label{External consistency}

External consistency refers to the alignment across the four key reporting and compliance mechanisms introduced by the Digital Services Act (DSA): (1) Transparency Reports, (2) the Statement of Reasons (SOR) Database, (3) Systemic Risk Assessment (SRA) Reports, and (4) Independent Audits. Each of these tools serves a different but interrelated role in assessing a platform’s efforts to moderate content, mitigate systemic risks, and fulfil legal obligations under the DSA.

Comparing Instagram’s Transparency Reports 2 and 3 with the SOR database reveals major discrepancies. For example, Transparency Report 3 records nearly 18 million account terminations, while the SOR database lists only 12.6 million for the same period—a difference of over 30\% \cite{meta_instagram_2024_2}. Similar gaps appear in Report 2. These findings point to problems with data integration and category definitions across reporting mechanisms.

\begin{table*}[ht]
\centering
\caption{Instagram Account Termination Metrics in DSA Transparency Reports [Table 15.1.c.(3) and (4)] 1–4 and Statement of Reasons (SoR) Database [Data accessed on the 3rd of July 2025].}
\label{tab:termination_metrics_updated}
\begin{tabular}{|l|r|r|r|}
\hline
\textbf{Transparency Report Period} & \textbf{Reported Terminations (TR)} & \textbf{Recorded in SoR} & \textbf{Difference (TR - SoR)} \\ \hline
Transparency Report 1  & 9,506,546 & 109,919 & 9,396,627 \\ \hline
Transparency Report 2  & 15,360,549 & 9,041,495 & 6,319,054  \\ \hline
Transparency Report 3  & 17,991,379 & 12,560,641 & 5,430,738  \\ \hline
Transparency Report 4  & 18,462,723 & 23,721,798 & $-$5,259,075 \\ \hline
\end{tabular}
\end{table*}

Instagram’s 2024 Systemic Risk Assessment (SRA) acknowledges the reporting fragmentation, noting that “reporting systems are not always fully aligned across internal metrics and DSA templates” and that some datasets submitted to the SOR are generated post hoc from logging systems rather than sourced directly from moderation workflows \cite[p. 33]{meta_sra_2024}. This introduces both definitional and temporal inconsistencies that hinder auditability and transparency.

These limitations are echoed in the 2024 Independent Audit conducted by Ernst \& Young. While the audit found that Meta’s moderation infrastructure and mitigation protocols were largely adequate, it flagged Instagram’s “asynchronous logging systems” as a key challenge to validating SOR completeness \cite[p. 18]{meta_audit_2024}. It further noted that Instagram “does not yet have a fully standardised process for reconciling discrepancies across transparency reports and SOR entries,” limiting the reliability of cross-mechanism verification.

Compounding these issues is the inconsistent use of terminology and categories between mechanisms. For example, SOR entries often reference `account restriction', while the transparency reports variably refer to `account terminations' or `service suspensions'. Platforms like TikTok also report account restrictions without clear disaggregation \cite{tiktok_tiktok_2024}, further complicating cross-platform comparisons. Although the DR attempts to harmonize language by linking transparency template fields with SOR keywords \cite[p. 8-11]{noauthor_annex_2024}, practical implementation remains uneven.

Moreover, the report identifies difficulties in harmonizing moderation outcomes across EU member states, stating that “risk indicators for political content over-enforcement rely on experimental classifiers that are not uniformly calibrated.” This lack of standardization limits the interoperability of data across transparency, SOR, and systemic risk reports and impairs external consistency evaluations.

Although the Delegated Regulation (DR) aims to reduce such inconsistencies by harmonizing reporting templates and aligning transparency fields with SOR keywords \cite[p. 8–11]{noauthor_annex_2024}, neither the SRA nor the audit confirms systematic use of these mappings. This raises concerns about the enforceability and comparability of DSA-mandated disclosures.

Beyond numerical inconsistencies, we also identify divergences in how the systemic risk mitigation measures are described and evaluated across the SRA and audit. For instance, the SRA highlights investments in safety-by-design and third-party partnerships as central mitigation strategies \cite[p. 6--8]{meta_sra_2024}, whereas the audit focuses more narrowly on governance and control frameworks, noting limited formalization and documentation of risk-tracking metrics across business units \cite[p. 10--11]{meta_audit_2024}. Furthermore, while the SRA positions automation as a cornerstone of risk mitigation (especially regarding harmful content and child safety), the audit raises concerns about the lack of transparency and interpretability in automated decision-making systems \cite[p. 19]{meta_audit_2024}. These strategic misalignments reflect broader tensions between risk framing, organizational accountability, and evidence-based compliance across mechanisms.

Besides, notable inconsistencies emerge in how illegal content and flagging mechanisms are represented across compliance documents. Instagram’s transparency reports provide limited granularity for illegal content removals, often aggregating categories such as “Other Illegal Content” without disclosing specific legal grounds or harmonized classification as required under Annex II of the Delegated Regulation (DR) and will likely change after the DR enters into force after August 2025 \cite[p. 7--11]{noauthor_annex_2024}. As a result, external evaluations under Article 15(1)(c) and 16 become more difficult.

These reporting inconsistencies are mirrored in Instagram’s 2024 Systemic Risk Assessment (SRA), which outlines systemic risks such as disinformation, gender-based violence, and civic discourse disruption, yet fails to align these categories with enforcement statistics from the SOR or transparency reports \cite[p. 10--15]{meta_sra_2024}. While the SRA emphasizes partnerships and safety-by-design, it does not quantify the efficacy of flagging mechanisms in mitigating specific risk vectors or link them to actionable thresholds for illegal content moderation. 

Taken together, the observed inconsistencies across transparency reports, SOR entries, the SRA, and the independent audit reveal a fragmented compliance landscape. The current implementation of external consistency mechanisms falls short of enabling a reliable evaluation of whether Instagram—or any VLOP—is fulfilling its DSA obligations. For external consistency to function as a meaningful compliance metric, platforms must not only synchronize reporting timelines and harmonize terminology but also establish interoperable technical pipelines that allow consistent, verifiable data exchange across all four DSA accountability layers.

\subsection{Historical Consistency}
\label{Historic consistency}

Historical consistency is examined by comparing trends across Instagram’s DSA transparency reports (1-4). The total volume of content removals dropped by more than 50\% between Report 1 and Report 2, by 67\% between Reports 2 and 3, and by almost 45\% between Reports 3 and 4 \cite{meta2023dsa1, meta2024dsa2, meta2024dsa3, meta2025dsa4}. These sharp fluctuations are not explained in the reports, making it difficult to interpret trends or assess data stability over time.

\begin{table*}[ht]
\centering
\caption{Percentage Change in Instagram's Policy Violations and Automated Removals Across Transparency Reports}
\label{tab:change_transparency}
\resizebox{\textwidth}{!}{%
\begin{tabular}{|p{4cm}|c|c|c|c|c|c|}
\hline
\textbf{Policy Violation} 
    & \textbf{Removed 1→2 (\%)} 
    & \textbf{Removed 2→3 (\%)} 
    & \textbf{Removed 3→4 (\%)} 
    & \textbf{Auto 1→2 (\%)} 
    & \textbf{Auto 2→3 (\%)} 
    & \textbf{Auto 3→4 (\%)} \\ \hline

Adult Nudity and Sexual Activity & +21.0 & +17.5 & -33.5 & +24.0 & +14.9 & -35.8 \\ \hline
Adult Sexual Solicitation \& Sexually Explicit Language & -- & -- & -- & -- & -- & -- \\ \hline
Bullying \& Harassment & +54.9 & -9.3 & -59.1 & +56.6 & -10.2 & -67.6 \\ \hline
Child Endangerment - Child Nudity and Physical Abuse & -52.1 & +490.1 & -- & -44.4 & -24.3 & -- \\ \hline
Child Endangerment - Child Sexual Exploitation & +141.8 & -24.5 & -- & +534.8 & -22.9 & -- \\ \hline
Child Sexual Exploitation, Abuse, and Nudity & -- & -- & -- & -- & -- & -- \\ \hline
Dangerous Organisation - Hate Orgs & +52.0 & -34.9 & -- & +75.9 & -36.3 & -- \\ \hline
Dangerous Organisation - Terrorism & +538.0 & -54.1 & -- & +603.7 & -56.3 & -- \\ \hline
Dangerous Individuals and Orgs & -- & -- & -- & -- & -- & -- \\ \hline
Fraud and Deception & -- & -- & -- & -- & -- & -- \\ \hline
Hate Speech & -0.8 & -3.9 & -50.6 & +2.9 & -4.3 & -52.0 \\ \hline
(Restricted Goods and Services) Drugs & -7.4 & -28.2 & -- & -7.9 & -45.5 & -- \\ \hline
(Restricted Goods and Services) Firearms & +376.7 & -65.8 & -- & +495.7 & -66.2 & -- \\ \hline
Restricted Goods and Services & -- & -- & -- & -- & -- & -- \\ \hline
Suicide and Self-Injury & -39.3 & +14.2 & -- & -34.9 & -8.1 & -- \\ \hline
Suicide, Self-Injury, and Eating Disorders & -- & -- & -- & -- & -- & -- \\ \hline
Spam & -- & -- & -37.6 & -- & -- & -36.2 \\ \hline
Third-Party IP Infringement & -- & -- & -- & -- & -- & -- \\ \hline
Cybersecurity & -- & -- & -- & -- & -- & -- \\ \hline
Violent and Graphic Content & +934.1 & +47.6 & -81.4 & +1176.2 & +42.9 & -82.9 \\ \hline
Violence and Incitement & +17.9 & -26.2 & -46.2 & +32.5 & -26.8 & -53.3 \\ \hline

\textbf{Total (including other violations)} 
    & -51.4 & -67.2 & -44.3 
    & -52.4 & -70.5 & -44.9 \\ \hline

\end{tabular}%
}
\end{table*}

For example, Report 2 introduced a new content category — `Spam' — that was absent in Report 1. While it accounted for the majority of removals in Report 2, its volume fell by over 80\% in Report 3 (see Table \ref{tab:instagram_combined_removals}). Categories like 'Firearms' (over 300\%) and 'Violent and Graphic Content' (over 900\%) exhibited increases  between Reports 1 and 2 (see Table \ref{tab:change_transparency}), which may relate to geopolitical events such as the Gaza conflict beginning in October 2023 \cite{mckernan_israel_2023}. However, the absence of explanatory notes for such surges diminishes interpretability.

Similarly, the 'Child Endangerment – Child Nudity and Physical Abuse' category saw a sixfold increase in removals between Reports 2 and 3, but a concurrent 24.3\% drop in automation use for the same category. This suggests increased human oversight but is not elaborated upon in the report narratives. While in Report 2 and 3 Meta states that it has around 40.000 moderators employed, it also states that it has "5.400 language agnostic reviewers globally" in Report 2
but only 2.500 in Report 3. Meta's last Transparency Report however indicates 4.600 language agnostic reviewers for their service in Report 4.

\begin{table*}[ht]
\centering
\caption{Yearly Overview of Organic Content Measures Including 'Other Violations' [Table 15.1.c.(1) in Instagram's Transparency Reports]}
\label{tab:yearly_report}
\resizebox{\textwidth}{!}{%
\begin{tabular}{@{}l p{4.5cm} r r r @{}}\toprule
\textbf{Report No.} & \textbf{Description} & \textbf{Total Numbers} & \textbf{Change from Previous Year} & \textbf{Difference} \\ \midrule
1. Report & First record of total violations & 76,298,413 & -- & -- \\
2. Report & Second record of total violations & 37,039,411 & Decrease & -39,259,002 (-51.44\%) \\
3. Report & Third record of total violations & 12,136,947 & Decrease & -24,902,464 (-67.25\%) \\
4. Report & Fourth record of total violations & 6,760,857 & Decrease & -5,376,090 (-44.30\%) \\ \bottomrule
\end{tabular}
}
\end{table*}

In sum, Instagram's reporting across the three consistency dimensions — internal, external, and historical — reflects challenges in both implementation and conceptual alignment with the DSA's compliance logic. These inconsistencies undermine transparency, hinder empirical oversight, and highlight the need for harmonized, standardized reporting practices as outlined in the Delegated Regulations.

\section{Challenges in Evaluating Consistency Across the SOR and Transparency Reports
}
\label{RQ3}

This section addresses RQ2 by examining the structural and semantic misalignments that undermine consistency checks across the DSA’s four core mechanisms: transparency reports, SOR entries, systemic risk assessments, and independent audits.

\subsection{Structural and Semantic Misalignments}
A primary challenge lies in the structural divergence between transparency reports and SOR entries, SRAs and audit reports. For example while transparency reports aim to provide aggregated, platform-level statistics (e.g., account terminations, content takedowns), SOR entries document individual moderation decisions. However, platforms often use divergent terminology and aggregation logics that make reconciliation difficult. For example, Instagram refers to "account terminations" in its transparency reports, while using "account restrictions" in SOR entries. The lack of harmonized taxonomies across mechanisms—despite efforts outlined in the Delegated Regulation's keyword mapping annex \cite[p. 8--11]{noauthor_annex_2024}—prevents automated cross-checking and undermines consistency.
Additionally, transparency reports often differ significantly in how platforms categorize ‘Terms of Service’ violations, which complicates comparisons across platforms. For example, as shown in Tables, \ref{tab:instagram_combined_removals}. Instagram's categories do not align directly with the standardized categories in the SOR database. This misalignment persists across platforms \cite{trujillo_dsa_2024}.

\subsection{Internal Inconsistencies in Quantitative Reporting}
Transparency reports frequently include category totals that do not align with underlying subtotals or violate Annex II's rules on double-counting and categorical separation \ref{tab:instagram_combined_removals}. For instance, Instagram’s “Total (Including Other Violations)” category combines disparate content types without clear attribution to specific contractual or legal grounds. Similarly, automated versus manual decision shares are inconsistently labelled and defined, complicating interpretability of compliance with Article 15(1)(c) DSA. Additionally, it makes historic comparisons less efficient if categories are renamed, merged or split up (see Table \ref{tab:instagram_combined_removals} and \cite{meta2025dsa4}).

\subsection{Lack of Alignment on Illegal Content and Flagging Mechanisms}

These challenges are reminiscent of issues seen in Germany's NetzDG and Austria's Communication Platform Law  \cite{noauthor_bundesgesetz_2020}. As Wagner et al. highlighted, differences in flagging structures for legal and contractual violations skewed transparency reports under NetzDG \cite{wagner_regulating_2020}. Facebook faced penalties for non-compliance with these standards, which underscores the importance of harmonising flagging and reporting mechanisms for more meaningful transparency mechanisms \cite{noauthor_germany_2019,bfj_press_agency_federal_2019}.
Similarly, the design of reporting mechanisms in the DSA (Art. 16) could face similar challenges influencing the transparency reporting quality and the SOR data \cite{sekwenz_it_2025}.
While Article 16 of the DSA mandates clear reporting of notices concerning illegal content, Instagram’s transparency reports group a large volume of removals under generic labels such as “Other Illegal Content”, without specifying whether these fall under national laws or contractual provisions \cite{meta_transparency_2024}. The SRA, meanwhile, focuses on high-level risks such as hate speech or child safety but does not quantify their treatment through user flagging or automated tools \cite[p. 16--18]{meta_sra_2024}. The audit corroborates these issues, noting that flagging processes lack documentation of how reports are classified, triaged, and attributed to legal grounds \cite[p. 22]{meta_audit_2024}. These gaps impair traceability and impede meaningful enforcement under Articles 16, 17, and 22.

\subsection{Misalignment of Categories Across Mechanisms}
Instagram's reporting allows for four categories of illegal content in its transparency reports  (see Table \ref{tab:instagram_illegal_expanded}), while its ‘Terms of Service’ categories include thirteen distinct violations and an ‘Other Violations’ category. This disparity complicates alignment with the SOR database, which groups ‘Illegal and Harmful Speech’ and other categories inconsistently across mechanisms and jurisdictions. For instance, only six SOR categories matched ‘Illegal and Harmful Speech’ in flagged contents, compared with the 1,090 entries for the same category in Instagram's Transparency Report 2.

\begin{table*}[ht]
\centering
\caption{Cross-Mechanism Traceability of Instagram’s Risk Mitigation and Reporting under the DSA}
\label{tab:risk_traceability}
\resizebox{\textwidth}{!}{%
\begin{tabular}{|p{4cm}|p{4.5cm}|p{4cm}|p{4cm}|}
\hline
\textbf{Systemic Risk (SRA)} & \textbf{Reported Mitigation (SRA)} & \textbf{Matching SOR Category} & \textbf{Matching Transparency Report Category} \\ \hline
Disinformation and Civic Discourse Harm & AI classifiers for deceptive content; partnerships with fact-checkers; downranking & “Scams and Fraud”, “Public Security Risk”, “Platform Service Abuse” & Fraud and Deception; Spam; Restricted Goods \\ \hline
Gender-based Violence & Safety-by-design measures; safety centre updates; user reporting channels & “Non-Consensual Behavior”, “Violence”, “Pornography or Sexualised Content” & Hate Speech; Bullying and Harassment; Adult Sexual Solicitation \\ \hline
Child Safety & Dedicated child safety classifiers; increased human oversight; third-party audits & “Child Protection”, “Nudity”, “Pornography or Sexualised Content” & Child Sexual Exploitation, Abuse, and Nudity; Suicide and Self-Injury \\ \hline
Hate Speech and Extremism & Risk forecasting tools; external policy partnerships; automated classifier updates & “Harmful or Illegal Speech”, “Public Security Risk” & Hate Speech; Dangerous Individuals and Orgs \\ \hline
Electoral Process Interference & Internal policy reviews; political ad transparency updates; partnership escalation protocols & [Not traceable in current SOR tags] & [No electoral or political content category] \\ \hline
Mental Health Harm (e.g. Eating Disorders) & Updated content guidelines; enhanced classifier precision & “Self-Harm”, “Violence”, “Pornography” & Suicide, Self-Injury, and Eating Disorders \\ \hline
\end{tabular}%
}
\end{table*}

Table~\ref{tab:risk_traceability} illustrates the difficulty of aligning systemic risks, mitigation actions, and reporting categories across Instagram’s DSA compliance mechanisms. Many risk areas identified in the Systemic Risk Assessment (SRA) are mapped inconsistently to Statement of Reasons (SOR) and Transparency Report categories. For instance, “disinformation” is linked to several different SOR and transparency categories, while some risks, such as electoral process interference, lack direct matches. This highlights significant gaps in cross-mechanism traceability and the need for harmonized, standardized classification across reporting instruments.

\subsection{Disjunctions Between SRAs and Transparency Reporting}
Systemic Risk Assessment Reports (SRA) and transparency reports are conceptually complementary: the former identify systemic risks (Art. 34) and mitigation strategies (Art. 35), while the latter document enforcement outcomes. However, we find no formal linkage between identified risk vectors (e.g., electoral disinformation, gender-based harms) and moderation outcomes reported by Instagram. The SRA identifies strategic mitigation pillars such as content classifiers and third-party partnerships \cite[p. 11--14]{meta_sra_2024}, yet these tools are not reflected in any disaggregated way in transparency reports. The absence of impact metrics renders it difficult to assess the effectiveness of mitigation and risks obfuscating accountability.
Instagram’s systemic risk assessment further illustrates this disjunction by listing multiple mitigation actions—such as “user education on deceptive content” and “enhanced fact-checking workflows”—without quantifying their impact on enforcement statistics or linking them to transparency report categories or SOR reasons. This disconnect reveals structural challenges in aligning risk mitigation strategies with operational transparency, weakening evaluability under the DSA.

For instance, Meta outlines initiatives to downrank “borderline content” that may propagate disinformation without violating platform rules. Yet, such interventions are neither reflected in moderation volumes (e.g. SOR) nor in risk quantification metrics (e.g. discussed in Table 15.1.c.(2) \cite{meta2025dsa4} under content demotion measures explicitly), illustrating the absence of traceability between systemic mitigation and platform enforcement outputs \cite[p. 8--9]{meta_sra_2024}.

\subsection{Ambiguities in Advertising Content}
Transparency reports often fail to distinguish between user-generated and advertising contents. For instance, Instagram reports combined figures for ‘business content measures’ across Facebook and Instagram \cite[p. 13]{meta2024dsa3}, making it unclear whether content removal data in Tables \ref{tab:instagram_combined_removals} and \ref{tab:change_transparency} include advertising violations. By contrast, TikTok explicitly differentiates between ‘advertising’ and ‘user’ policies in its Transparency Report 2  \cite{tiktok_dsa_2023,tiktok_dsa_2024}. Standardizing these distinctions could improve cross-platform comparisons and enable deeper insight into content moderation practices.

\subsection{Lack of Standardized Measures of Automated Accuracy}
Platforms report accuracy of automated moderation differently, further complicating evaluations of consistency and measuring compliance. Instagram reports an ‘automation overturn rate’ of 8.4\%, whereas TikTok provides ‘specific accuracy’  (98.6\%) and ‘error rate’  (1.4\%) metrics. Without standardized definitions for accuracy measures, cross-platform comparisons lack reliability and consistency.

\paragraph{We have identified and evaluated seven key factors – structural and semantic misalignments, internal inconsistencies, ambiguous categorization of illegal content, disjunctions between systemic risk assessments and transparency reporting, advertising content ambiguities, inconsistent automation reporting standards, and misaligned categorization across mechanisms – because these issues directly affect the reliability, comparability, and auditability of platform compliance data under the DSA. These factors were prioritized due to their significant impact on regulatory coherence and meaningful oversight, as well as their prevalence in existing compliance reporting practices. An overarching theme across these challenges is the critical need for consistency, both conceptually and operationally. The issues identified frequently amplify each other, undermining the mechanisms introduced by the DSA and complicating effective regulatory interventions. Conversely, addressing these interconnected issues in a structured way can enhance trust, facilitate regulatory accountability, and ultimately reinforce platform governance. In the following section, we present targeted recommendations and practical solutions aimed at systematically improving consistency across these mechanisms, thus bridging the identified challenges to a more robust and empirically verifiable compliance framework.}

\section{Recommendations to Evaluate Consistency Across DSA Compliance Mechanisms}
\label{RQ4}

This section addresses the third research question: \emph{How can a consistency-based approach support effective oversight and auditing under the DSA?} Building on the three-level consistency framework presented in Section \ref{RQ2}, we propose opportunities to enhance the regulatory effectiveness of DSA reporting mechanisms through greater alignment and mutual reinforcement across tools.

\textbf{Current Gaps in Compliance Evaluations}
European Commission investigations into platforms such as Instagram, TikTok, X, and AliExpress have not yet disclosed specific infringements under Articles 15, 17, 24, or 42 of the DSA. However, our comparative analysis in Section \ref{RQ2} reveals systemic inconsistencies across the transparency reports, the SOR database, Instagram’s systemic risk assessment (SRA), and its 2024 independent audit. These gaps call for unified evaluation structures that allow regulators to identify risks, assess reporting quality, and support compliance enforcement with more precision.

\textbf{Integrating Multiple Consistency Checks}
The three levels of consistency—internal, external, and historical—can be extended to enable richer evaluations across multiple DSA tools. In particular, we recommend:

1. \emph{Cross-Referencing SORs and SRAs}: Platforms should be required to explicitly reference consistency findings from the SOR database in their systemic risk assessments. For example, discrepancies in moderation reasons or decision volumes could be flagged as a systemic governance risk. Instagram’s 2024 SRA partially implements this but lacks specific quantitative reconciliation.
Although Instagram’s risk report concedes that “transparency reporting discrepancies may reflect systemic governance risks,” it stops short of integrating these observations into a structured consistency analysis or remediation workflow \cite[p. 15]{meta_sra_2024}. Future systemic risk assessments should be required to flag anomalies in DSA reporting—such as divergent moderation volumes or absent removal justifications—as traceable risks under Articles 34 and 35 DSA.

2. \emph{Aligning Audit and Reporting Metrics}: External audits should validate the quantitative consistency of transparency reports and SORs as part of their control assessment. While EY’s audit of Instagram mentions SOR data validation, it stops short of conducting full reconciliation checks. Future audits should be standardized to include consistency assessments that also disclose information of the auditing time frame, scope, data, and methodology used in greater detail, e.g. disclosing information about sampling criteria \cite{sekwenz_doing_2025}.

3. \emph{Creating a Unified Compliance Matrix}: A structured matrix that links transparency report metrics, SOR entries, SRA findings, and audit conclusions would improve traceability and support empirical accountability reviews. Such a matrix could be integrated into Annex templates provided by the Delegated Regulation (DR) or as stand-alone section in SRAs and audits.

\textbf{Standardization Priorities}

\emph{Terminology and Classification Alignment}: Shared definitions for moderation outcomes (e.g., account restriction vs. termination) must be enforced across all DSA mechanisms. The DR’s keyword mappings (\cite[p.\ 8--11]{noauthor_annex_2024}) should become mandatory and used accross reporting mechanisms.

\emph{Synchronized Reporting Timeframes}: The DR already introduces biannual reporting cycles (Article 2), but full harmonization of report publication dates is necessary to allow meaningful cross-mechanism analysis. To prohibit for example that moderation spikes linked to elections are covered by the Transparency Report of one platform, but not of another.

\emph{Automation Reporting Standards}: As automated systems remain central to moderation, we echo the DR’s call for detailed classifier reporting, including accuracy, precision, and recall (\cite[p.25]{noauthor_annex_2024}). SRAs and audits should explicitly assess the performance of such systems using standardized indicators. Additionally, it is essential to show where accuracy is achieved. This includides information about content types (reals, live content, video, comments, images, generative Artificial Intelligence content, etc.) and platform's policies (terrorist content, disinformation, hate speech, etc.). 

Meta’s own systemic risk assessment echoes this need, stating that “without harmonized metrics for automation accuracy and over-removal risk, comparative evaluations across time or region are not feasible” \cite[p.~10]{meta_sra_2024}. This admission underlines the importance of embedding standardized metadata schemas and accuracy indicators—such as precision, recall, and classifier confidence thresholds—across all reporting instruments as foreseen in the Delegated Regulation \cite[p.~25]{noauthor_annex_2024}.

By operationalization consistency across multiple layers of the DSA’s compliance regime, we offer a roadmap for more reliable, evidence-based assessments of platform governance. The ability to triangulate insights from different DSA tools supports both proactive risk mitigation and reactive enforcement.

\section{Leveraging Consistency for Regulatory Evaluation}
\label{findings}

The central message of our article is simple but crucial: \textit{consistency matters} in platform governance under the DSA. Our proposed consistency-based methodology enables multi-layered insights that go beyond surface-level reporting metrics. Evaluating different forms of consistency across four key DSA mechanisms—transparency reports, the Statement of Reasons (SOR) database, systemic risk assessment (SRA) reports, and independent audits—provides a more holistic understanding of how platforms comply with their obligations.

\section{Discussion}
\label{Discussion}

In this article, we analysed the transparency and accountability mechanisms introduced by the Digital Services Act (DSA), focusing on Instagram as a case study. By combining legal analysis with a methodological framework for evaluating reporting consistency, we offer a novel perspective on how platforms disclose moderation activities and systemic risks under the DSA.

Addressing the complexities of DSA implementation and compliance requires close collaboration between computer science and law. As this article demonstrates, robust evaluation of platform reporting mechanisms is only possible by combining technical expertise in data analysis, information systems, and algorithmic transparency with legal knowledge of regulatory requirements, fundamental rights, and enforcement processes. Regulators, auditors, and researchers each bring vital perspectives: legal scholars interpret obligations and rights, computer scientists operationalize empirical consistency checks, and auditors synthesize evidence to inform oversight. Interdisciplinary and cross-sectoral engagement is essential, as the task of making DSA mechanisms effective is a collective endeavor—one that depends on ongoing dialogue, shared learning, and the integration of diverse skills and backgrounds to navigate this evolving regulatory landscape.

\textbf{Summary of Contributions.}
We introduced a structured framework for assessing four levels of consistency—internal, external, historical, and cross-mechanism—across four key DSA-mandated mechanisms: Transparency Reports, Statement of Reasons (SOR) entries, Systemic Risk Assessment (SRA), and Independent Audits. These mechanisms were evaluated through three primary lenses: legal compliance, practical use case analysis, reporting obstacles, and opportunities for harmonization and improved compliance.

Our legal-doctrinal analysis clarified how Articles 15, 16, 17, 24, 34, 37, and 42 of the DSA, together with the Delegated Regulation, define transparency obligations. Our methodological approach demonstrated how internal inconsistencies in Instagram’s reports (e.g., `Other Violations' lumping), external misalignment with SOR data, and lack of historical coherence (e.g., sudden category changes like `Spam') compromise the value of transparency in practice.

\textbf{Challenges and Limitations.}
Despite the promise of transparency tools under the DSA, several issues remain unresolved. Disparities in reporting formats, vague or undefined categories, and asynchronous reporting cycles undermine comparability across platforms and even across mechanisms within the same platform. As shown in Section \ref{External consistency}, Meta’s SRA and audit report both acknowledge that SOR submissions are often derived post-hoc, raising concerns about the accuracy and origin of reported data.

As demonstrated in Section \ref{RQ2}, audit and SRA reports are not currently standardized, hindering effective cross-mechanism validation. These issues point to an urgent need for more integrative and standardized governance instruments.

\textbf{Opportunities for Enhancing Compliance Evaluation.}
As discussed in Section \ref{RQ4}, combining insights from transparency reports, SORs, SRAs, and audit reports offers a more holistic approach to evaluating DSA compliance. We argue that the proposed consistency layers enable a more rigorous and triangulated evaluation strategy (if reporting and auditing is standardized in a meaningful way). Aligning definitions, synchronizing timelines, and mandating shared classification schemes could turn these disparate mechanisms into a coherent reporting ecosystem.

This approach also strengthens the evidence base for regulatory interventions. For example, discrepancies in account restriction figures across mechanisms could be flagged for audit, while sudden shifts in moderation categories across time may suggest procedural inconsistencies warranting further investigation.

\textbf{Interlocking Potential of DSA Mechanisms.}
The integration of SRAs and independent audits into the DSA ecosystem is not merely a compliance requirement—it is a strategic opportunity. These instruments can act as checks on platform self-reporting and provide regulators with contextual insights that go beyond raw numbers. When triangulated with SOR data and transparency reports, they enable a systems-level evaluation of risk, and procedural consistency.

In future iterations of the DR templates, Guidelines, and best practices policymakers could consider requiring platforms to explicitly reference and reconcile findings across these mechanisms. Doing so would elevate the status of consistency not just as a desirable feature, but as a formal legal expectation.

\textbf{Balancing Standardization and Innovation.}
While the need for shared metrics and harmonized terminology is clear, regulators must avoid over-constraining platform innovation—especially with regard to platform-specific Terms and Conditions (TaC). As our analysis shows, standardization should focus on ensuring traceability and comparability, not imposing a one-size-fits-all model.

In sum, we argue that consistency is not just a metric for reporting quality but a proxy for the DSA’s regulatory effectiveness. By leveraging all four transparency mechanisms in tandem, we unlock a more actionable and robust understanding of platform compliance.

\subsection{Limitations}
\label{Limitation}

This study has several limitations stemming from the novelty of the DSA framework and the evolving nature of its implementation. First, the misalignment of reporting time frames restricts full comparability across mechanisms. While Instagram’s second and third transparency reports (1 October 2023–31 March 2024 and 1 April 2024–30 September 2024, respectively) overlap with the launch of the DSA Transparency Database (September 2023), the first report does not. As a result, historical consistency across all three reports cannot be triangulated with SOR data for the initial period \cite{noauthor_statements_nodate, noauthor_commission_2023}.

Second, the overall utility of transparency mechanisms remains constrained by definitional ambiguities. Core concepts—such as “account restriction,” “violation categories,” and “automation overturn rate”—are not standardized across transparency reports, SORs, systemic risk assessment (SRA) reports, and audit statements. This lack of shared terminology limits the ability to conduct fully integrated evaluations across mechanisms.

Third, both the SRA and audit reports analyzed in this study follow non-standardized formats and diverge significantly in the level of detail provided. For example, while the audit document by Ernst \& Young acknowledges the asynchronous nature of SOR logging, it does not quantify the potential reporting gap or specify corrective measures. Similarly, Instagram’s SRA report describes internal compliance frameworks but does not provide sufficient granularity for meaningful external validation. These gaps limit the strength of cross-mechanism consistency checks.

Lastly, the Delegated Regulation (DR) under Articles 24 and 88 of the DSA was only recently adopted and its harmonized reporting templates will not become mandatory until 1 July 2025. Therefore, this study captures a transitional reporting phase. As future reports begin to reflect DR-aligned structures—including standardized templates, keyword mappings, and clearer definitions—comparability and reliability of compliance evaluations are expected to improve. Our findings provide an early diagnostic of transparency under the DSA. They do not represent a definitive compliance audit.

\section{Conclusion}
\label{Conclusion}

Transparency mechanisms under the Digital Services Act (DSA) mark a significant regulatory advance in improving accountability in platform governance. Yet, as this study has shown, their effectiveness depends not merely on disclosure, but on \emph{consistency}—across reporting structures, categories, time frames, and platforms. We argue that transparency becomes meaningful only when reporting mechanisms such as the Transparency Reports, Statements of Reason (SORs), Systemic Risk Assessment (SRA) reports, and Independent Audits are not only publicly available but also mutually verifiable.

Our findings demonstrate that evaluating consistency across multiple levels—internal, external, historical, and cross-platform—can reveal critical discrepancies and structural gaps in reporting. These inconsistencies highlight both the challenges of implementing the DSA in its early stages and the opportunities to develop a more robust compliance ecosystem. We identified ambiguities in reporting illegal content, mismatches in moderation terminology, misaligned time frames, and unclear accuracy metrics, all of which hinder comparative evaluation and undermine regulatory enforcement.

By analyzing Instagram as a case study, we proposed a novel consistency-based framework for assessing DSA compliance. This framework combines legal analysis with empirical validation across complementary reporting mechanisms. In doing so, it offers a pathway for future research, regulatory development, and audit practice.

Looking ahead, the full potential of the DSA regime will depend on the adoption of harmonized standards, such as those outlined in the Delegated Regulation (DR). Aligning terminology, reporting cycles, and evaluation criteria will enable regulators, researchers, and platforms to unlock the transformative capacity of of the regulation — not merely as a formal obligation, but as a substantive mechanism for accountability, comparability, and democratic oversight in digital environments.

%\bibliographystyle{ACM-Reference-Format}

\pagebreak

\section{Appendices}

%%If your work needs an appendix, add it before the
%%``\verb|\end{document}|'' command at the conclusion of your source
%%document.

%%Start the appendix with the ``\verb|appendix|'' command:
%\begin{verbatim}

\appendix
\section{Overview of Key Legal References and Provisions}
\begin{table}[ht]
\small
\centering
\caption{Key Legal Provisions Referenced in this Article}
\label{app:legal-table}
\begin{tabularx}{\textwidth}{|p{2.7cm}|p{2cm}|X|}
\hline
\textbf{Provision} & \textbf{Jurisdiction} & \textbf{Summary and Relevance} \\ \hline
NetzDG (2017) & Germany & Requires semi-annual transparency reports on content moderation; criticized for non-standardized categories and incomplete reporting. Fines for non-compliance. \\ \hline
KoPl-G (2020) & Austria & Mandates transparency reporting for communication platforms; modeled after NetzDG, with similar shortcomings. \\ \hline
DSA Article 14 & EU & Requires platforms to provide clear terms and conditions; anticipates a centralised TaC Database. \\ \hline
DSA Article 15 & EU & Sets general transparency requirements for intermediary services, including machine-readable, public reports. \\ \hline
DSA Article 16 & EU & Requires notice-and-action mechanisms for reporting illegal content. \\ \hline
DSA Article 17 & EU & Establishes the Statement of Reasons (SOR) mechanism: platforms must justify each moderation action, referencing legal or contractual grounds. \\ \hline
DSA Article 24 & EU & Extends transparency requirements to online platforms. \\ \hline
DSA Article 34 & EU & Introduces periodic systemic risk assessments, focusing on fundamental rights and public interest risks. \\ \hline
DSA Article 37 & EU & Mandates independent audits of VLOP/VLOSE risk assessments and reporting mechanisms. \\ \hline
DSA Article 42 & EU & Sets biannual reporting requirements for VLOPs and VLOSEs. \\ \hline
DSA Delegated Regulation for transparency reporting & EU & Lays down harmonized templates, taxonomies, and metadata for transparency reports, ensuring comparability and standardization across platforms (Arts. 15(3), 24(2), 42 DSA). \\ \hline
DSA Delegated Regulation for conducting audits & EU & Specifies requirements for conducting audits and systemic risk assessments under the DSA. \\ \hline
Charter of Fundamental Rights (CFR) & EU & Referenced in DSA Article 34(1)(b) as the basis for risk assessment focus on fundamental rights. \\ \hline
\end{tabularx}
\end{table}

%\end{verbatim}


%%\appendix

\end{document}